# THE TOO-LATE-CHOICE EXPERIMENT: BELL'S PROOF WITHIN A SETTING WHERE THE NONLOCAL EFFECT'S TARGET IS AN EARLIER EVENT

Avshalom C. Elitzur[1], Eliahu Cohen[1,2], Tomer Shushi[3]

[1]*Iyar*, The Israeli Institute for Advanced Research, Rehovot, Israel.
avshalom@iyar.org.il
[2]H.H. Wills Physics Laboratory, University of Bristol, Tyndall Avenue, Bristol, BS8 1TL, U.K..
eliahu.cohen@bristol.ac.uk
[3]University of Haifa, Haifa 27019, Israel
tomer.shushi@gmail.com

*In the EPR experiment, each measurement addresses the question "What spin value has this particle along this orientation?" The outcome then proves that the spin value has been affected by the distant experimenter's choice of spin orientation. We propose a new setting where the question is reversed: "What is the orientation along which this particle has this spin value?" It turns out that the orientation is similarly subject to nonlocal effects. To enable the reversal, each particle's interaction with a beam-splitter at $t_1$ leaves its spin orientation superposed. Then at $t_2$, the experimenter selects an "up" or "down" spin value for this yet-undefined orientation. Only after the two particles undergo this procedure, the two measurements are completed, each particle having its spin value along a definite orientation. By Bell's theorem, it is now the "choice" of orientation that must be nonlocally transmitted between the particles upon completing the measurement. This choice, however, has preceded the experimenter's selection. This seems to lend support for the time-symmetric interpretations of QM, where retrocausality plays a significant role. We conclude with a brief comparison between these interpretations and their traditional alternatives, Copenhagen, Bohmian mechanics and the Many Worlds Interpretation.*

Quantum nonlocality [1], rigorously proven by the Bell [2] and GHZ [3] theorems, has a few temporal counterparts, where the quantum effects appear to go in the backwards time direction [4]. However, unlike non-locality, however, retrocausality was never considered such a pivotal issue, let alone requiring rigorous proofs. This neglect is inappropriate, because some modern interpretations of quantum mechanics, mainly Cramer's [5] transactional interpretation (TI) and Aharonov's [6] two state-vector formalism (TSVF), render time-symmetric causality essential for better understanding QM. In the EPR case, for example, the measurement's effect simply zigzags along the two particles' world-lines, thereby what appears nonlocal in *space* turns out to be perfectly local in *spacetime* [7]. Other quantum peculiarities, such as the uncertainty principle and the measurement problem, may similarly benefit from this perspective [7,8]. TSVF further derives several surprising predictions [6], of which many have already won empirical confirmation.

In this article we offer a novel and simple demonstration of quantum retrocausality. We begin in Sec. 1 with a non-exhaustive review of quantum retrocausality. The essential features of the standard EPR experiment are then briefly recounted in 2, followed in 3 and 4 by our "Too-Late Choice" variant. Sec. 5 shows how spin value, rather than spin direction, can be arbitrarily chosen. Sec. 6 describes the unique entanglement resulting from this reversal of measurement stages. Sec. 7 utilizes a modification of the "Too-Late Choice" experiment and demonstrates its straining the alternative account of Bohmian mechanics. Sec. 8 concludes with a brief comparison between the ways different schools of interpretation respond to the challenge of apparent quantum retrocausality.

## 1. Can A Quantum Effect Precede its Cause?

We begin with a simple working definition. Let "Orthocausality" denote the ordinary temporal direction of causation, which means that quantum measurement determines the particle's state at that moment as well as the consecutive ones until the next measurement. "Retrocausality," then, asserts that the measurement determines also the particle's *past*, back to the preceding measurement.[1] Following are some arguments in favor of the latter.

The foundations of time-symmetry in QM have been laid out by the work of Aharonov, Bergmann and Lebowitz (ABL), who used a symmetric construction of statistical ensembles (now known as pre- and post-selection) to reveal a time-symmetric probability distribution of measurement outcomes [9]. This finding was later generalized into a comprehensive interpretation of QM, namely the Two-State-Vector Formalism (TSVF). This school then predicted some surprising effects that, although perfectly consistent with standard quantum theory, emerged only within the TSVF framework.

The first experiment suggesting quantum retrocausality was Wheeler's "delayed choice" gedankenexperiment [10], later realized in [11]. In that setting, the experimenter's last-minute decision is argued to determine whether a photon coming from a very far source has come, *all along*, as a wave or a particle. Yet Wheeler himself did not conclude that the experimenter's decision really determines the particle's past. On the contrary, following the Copenhagen school, he preferred to

---

[1] The two accounts are not equivalent. Orthocausality is often presented as exclusive, admitting only one time direction. Retrocausality, in contrast, always comes within a time-symmetric framework.

*undermine the notion of reality in all times* ("No phenomenon is a phenomenon until it is an observed phenomenon" [10]).

Elitzur, Dolev & Zeilinger [12] were more explicit in their analysis of a time-reversed EPR ("RPE"). Let two excited atoms emit single photons towards the same beam-splitter. Suppose only one photon in detected, with no indication which of the two atoms emitted it. This uncertainty induces genuine superposition on the two atoms' states, moreover entangling them into a full EPR-Bell state. Here a temporal novelty emerges: The entangling event (*i.e.*, the detector's click) seems to reside in the two atoms' *future* rather than past. Elitzur *et al.* also pointed out the "delayed choice" aspect in this setting: By choosing at the last moment whether or not to insert a BS where the two possible photon paths cross, the experimenter seems to be deciding whether or not the two atoms *have been entangled*.

This setting was later used by Elitzur & Dolev [13] for their Quantum Liar experiment, where the retrocausal argument was straightforwardly based on Bell's proof. Here too, a single photon is emitted from one out of two atoms, its origin remaining indeterminate. This again entangles them into an EPR-Bell pair. On this pair, they employed the standard Bell setting where each atom undergoes a spin measurement along one of three orientations. There was, however, one crucial modification: One of these three spin measurements was replaced with a different measurement that amounts to asking, "*Is this atom entangled with the other one?*" In half of the cases, the outcome must be "No." Yet this outcome too, just like those of the two spin measurements, is imposed by entanglement, catching Nature, so to speak, blushing in self-contradiction. Several comments on this gedankenexperiment, *e.g.* [14-16], especially Cramer's analysis [17] and Kastner's "possibilist" version of TI [18], indicate that the application of Bell's theorem makes retrocausality merit serious consideration.

Elitzur & Cohen [19] reviewed these retrocausal approaches as well as several measurement methods, such as weak [20] and partial measurements [21], which they called "incomplete." They also proposed gedanken combinations of them to better elucidate quantum retrocausality.

Aharonov, Cohen & Elitzur [7] offered another twist to the EPR-Bell setting with the addition of weak measurements preceding the final strong ones. Admittedly, the predicted results can be explained by orthocausality, *i.e.*, that the weak measurements have inserted some slight bias into the particle pair, which later affected the final strong measurements. Yet Aharonov *et al*. argued that the more natural account is that of the TSVF, where the backwards evolving (post-selected) states determined the outcomes of the earlier weak measurements.

More recently, Elitzur and Cohen studied a simple quantum interaction [22] that ends up with an event and a nonevent, together appearing to violate momentum conservation. They showed that the nonevent in question is due to a "Quantum Oblivion" effect, where a very brief virtual interaction undergoes "unhappening." Oblivion, they argued, underlies quantum IFM, erasure and several other peculiar effects. Venturing further to theory, they proposed [22, 23] a retrocausal evolution that accounts for such self-cancellation, involving exchange of negative physical values between earlier and later events. These works offer a comprehensive framework for the study of both weak and strong measurements, with several intermediate phenomena that call for further study.

This is the background for the argument proposed below. We gedankenly apply Bell's theorem to an EPR-Bell setting where the spin measurement's two stages, namely, *i*) the choice of spin orientation to be measured and then *ii*) the final outcome thereof, are made in reversed order. Bell's proof in this case seems to inject the nonlocal influence into the particles' past.

2. **The EPR Experiment: "What is the Particle's Spin Value along this Orientation?"**

Recall first the standard EPR-Bell setting (Fig. 1).

I. A particle pair is prepared in the singlet state $\frac{1}{\sqrt{2}}\left(|\uparrow\rangle_A|\downarrow\rangle_B - |\downarrow\rangle_A|\uparrow\rangle_B\right)$. Each particle (A/B) travels far away from the other (Fig. 1).

II. Then, for each particle, one out of three possible spin orientations, *e.g.*, from the co-planer set {$\alpha/\beta/\gamma$}, is randomly chosen to be measured. Temporally speaking, this is the spin measurement's *preparation*, by the particle's entrance into the appropriately aligned Stern-Gerlach (SG) magnet.

*III.* Measurement is then completed by detectors placed at each SGM's two exits. Upon a particle's detection by one of these detectors, a spin value (↑/↓) is obtained along the chosen orientation.

*IV.* Over many times, correlations between the ↑/↓ values are expected within the A/B pairs. By Bell's theorem, each pair of spin values turns out to depend on the two orientations {*α,β,γ*}, chosen an instant earlier.

*V.* Local realism may naively account for these correlations as pre-existing spin values, specified in advance for each possible orientation, to be only passively revealed by any future measurements.

*VI.* However, Bell's theorem proves[2] that no such pre-existing values can be accommodated with the experimenters' simultaneous and independent free choices of spin orientations. It is therefore these choices, plus their random outcomes, that non-locally affected the distant particle's spin value.

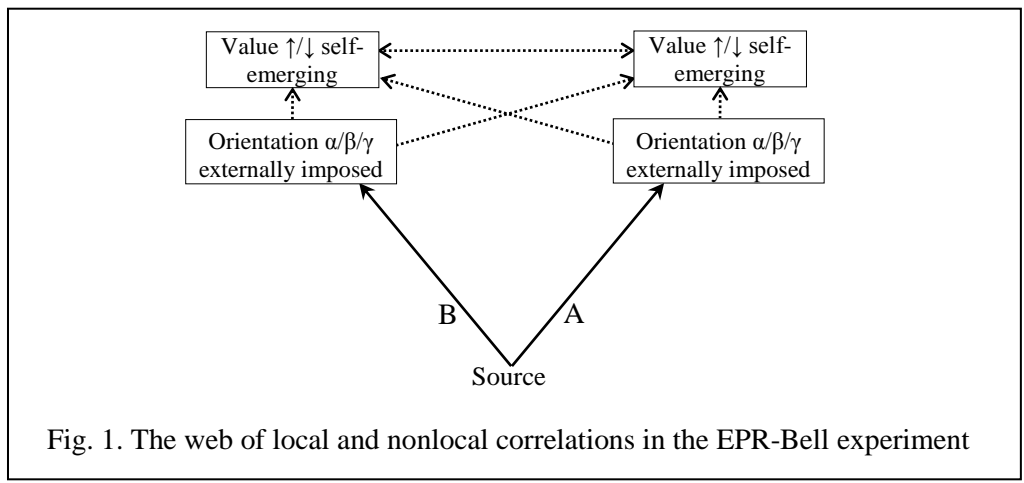

Fig. 1. The web of local and nonlocal correlations in the EPR-Bell experiment

### 3. A Slight Modification: Leave the Particle to "Choose" the Spin-Orientation to be Measured

To introduce our Too-Late version of the EPR, let us reconsider the measurement in the standard setting (stages 2*II*-2*III* above). Each particle's spin value turns out to be determined by *i)* the spin orientation measured on the other particle – freely chosen at the last moment by the distant experimenter – and *ii)* that measurement's outcome – randomly imposed by quantum uncertainty (Fig. 1). On this combination Bell's proof rests.

---

[2] Assuming orthocausality, to be addressed below.

Suppose however that the choice of spin-orientation is also left to the particle: A simple three-port beam-splitter sends it to one out of three SGMs, aligned along *α*, *β*, or *γ*. Only upon the final detection, then, will the particle "decide" which choice it was.

Does this modification make the non-locality proof weaker? Apparently, giving up the experimenter's free decision amounts to losing the outcome's "external" component, to be nonlocally transmitted between the particles. A closer look, however, shows that this component does not have to be external. First, the particles are not correlated for position and momentum. Moreover, the three positions of the SGMs can be arbitrarily placed for each BS. Therefore, *the particles cannot "conspire" in advance about their choices of spin orientation*. Hence they still have to inform one another, *non-locally*, about the spin orientation each particle eventually "chooses."

This simplification, therefore, leaves the non-locality proof intact while enabling the temporal exchange between the spin value and spin orientation outcomes.

## 4. The Too-Late-Choice Experiment: "Along which Orientation does the Particle have this Spin Value?"

Suppose, then, that we could somehow first obtain a certain spin value for each EPR particle and only then subject the pair to a question like "both spins are ↑, so to which orientations does this ↑ pertain for A and for B?" Then, it would be these orientations that Bell's theorem obliges to be non-locally affected.

Here, then, is the Too-Late Choice experiment (Fig. 2)

I. A particle pair is prepared as in (2*I*).

II. For each particle, the spin orientation (2*II*) is left undecided as the particle interacts with a three-port beam-splitter which makes its momentum superposed ($p_1$/$p_2$/$p_3$) towards the three SG magnets pre-aligned along (*α*/*β*/*γ*). The particle thus goes through all three magnets superposed.

III. With a special delicate measurement (see Sec. 5), each particle is measured only for its spin *value* (↑/↓), the orientation still left superposed.

IV. Only then is measurement completed to reveal the orientation to which this value pertains.

*V.*     Nonlocal correlations are now expected between the spin orientations, just as they were with spin values in (2*IV*).

*VI.*     Local realism may try to preserve the temporal order in the following way: The earlier choice of spin orientation, although not yet known to the observer, has already been decided by the particles earlier, then transmitted non-locally *at that time*. Therefore, the spin value obtained at (2*IV*) was by then pertaining to this unknown-yet-definite orientation.

*VII.*     Here, however, Bell's inequality would not suffice to rule out the orthocausal alternative. Another falsification is given below by showing, through quantum interference, that the spin orientation was genuinely superposed even after spin value was determined.

*VIII.*     It is therefore the momentum pairs of $p_1/p_2/p_3$, otherwise uncorrelated, and for which *any choice seems to be too late*, which become non-locally correlated by virtue of the later, familiar correlations between the spin values.

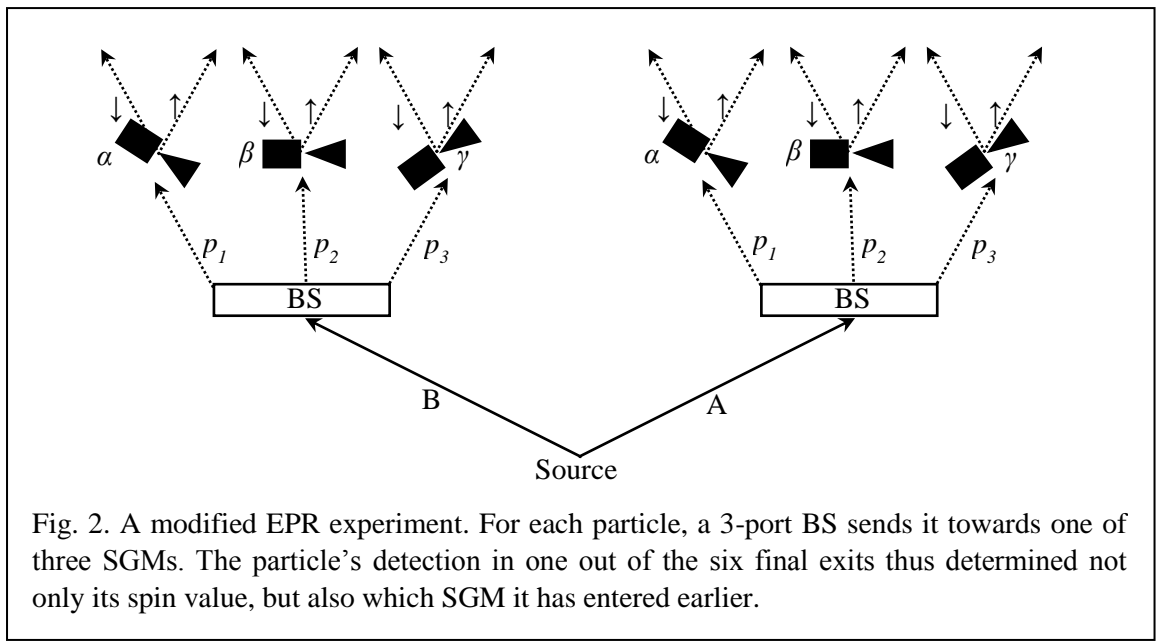

Fig. 2. A modified EPR experiment. For each particle, a 3-port BS sends it towards one of three SGMs. The particle's detection in one out of the six final exits thus determined not only its spin value, but also which SGM it has entered earlier.

### 5. But How can the Spin Value be Arbitrarily Chosen?

Of course, the above "somehow" must be specified in order to make the experiment feasible. How can a measurement give a definite spin value for all orientations?

Quantum superposition can do just that (Fig. 3). First let the two EPR particles going through the above setting be heavy atoms. Now delay each atom within its six possible exits ($\alpha/\beta/\gamma$)·($\uparrow/\downarrow$). Next send a photon along the three "up" exits and another photon through the three "down"s. As only one photon emerge unabsorbed, we know that the particle's spin is "up" or "down" without yet determining the orientation.

Finally, complete the measurement: Place detectors on the three "up/down" SGM exits where the atom must reside. A click will occur in one of them, giving the spin orientation to which the value pertains.

With the two particles' spin values thus fixed earlier, it is the orientations which must nonlocally obey Bell. Consider the simplest case, *e.g.*, $\uparrow_A,\uparrow_B$. An anti-correlation is immediately expected: If A's "up" is found to be along, say, $\alpha$, then B must *not* be found in $\alpha$.

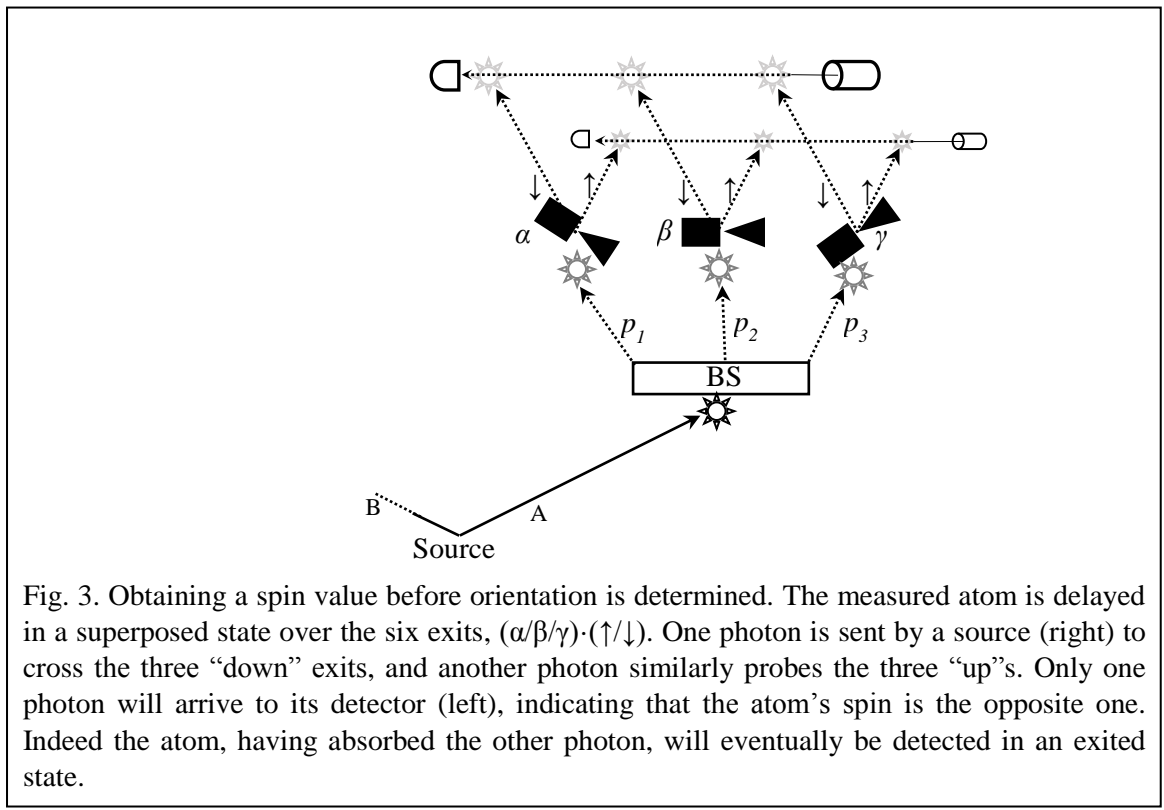

Fig. 3. Obtaining a spin value before orientation is determined. The measured atom is delayed in a superposed state over the six exits, ($\alpha/\beta/\gamma$)·($\uparrow/\downarrow$). One photon is sent by a source (right) to cross the three "down" exits, and another photon similarly probes the three "up"s. Only one photon will arrive to its detector (left), indicating that the atom's spin is the opposite one. Indeed the atom, having absorbed the other photon, will eventually be detected in an exited state.

## 6. The Too-Late Entanglement

The above setting presents a new variation on the EPR-Bell experiment. Instead of the superposition of spin *values* in the standard EPR-Bell, here the spin *orientations* are superposed prior to the final measurement. If, for instance, the spin value of particle A is fixed to be "up", its state would be

$$|\psi^A\rangle_\uparrow = \frac{1}{\sqrt{3}}\left(|\alpha\rangle_\uparrow + |\beta\rangle_\uparrow + |\gamma\rangle_\uparrow\right), \qquad (1)$$

and similarly for the B particle. However, fixing an "up" value for the two of them, must involve nonlocal correlations that exclude the three same-direction cases

$$|\psi^{AB}\rangle_{\uparrow\uparrow} = \frac{1}{\sqrt{6}}\left[|\psi^A\rangle_\uparrow |\psi^B\rangle_\uparrow - \frac{1}{3}\left(|\alpha^A\rangle_\uparrow |\alpha^B\rangle_\uparrow + |\beta^A\rangle_\uparrow |\beta^B\rangle_\uparrow + |\gamma^C\rangle_\uparrow |\gamma^C\rangle_\uparrow\right)\right]. \qquad (2)$$

Taking into account all combinations, we find the entangled state

$$|\text{TooLate}\rangle = \frac{1}{\sqrt{30}}\left[|\psi^A\rangle_\uparrow |\psi^B\rangle_\uparrow + |\psi^A\rangle_\uparrow |\psi^B\rangle_\downarrow + |\psi^A\rangle_\downarrow |\psi^B\rangle_\uparrow + |\psi^A\rangle_\downarrow |\psi^B\rangle_\downarrow - \right.$$
$$\left. \frac{1}{3}\left(|\alpha^A\rangle_\uparrow |\alpha^B\rangle_\uparrow + |\beta^A\rangle_\uparrow |\beta^B\rangle_\uparrow + |\gamma^C\rangle_\uparrow |\gamma^C\rangle_\uparrow + |\alpha^A\rangle_\downarrow |\alpha^B\rangle_\downarrow + |\beta^A\rangle_\downarrow |\beta^B\rangle_\downarrow + |\gamma^C\rangle_\downarrow |\gamma^C\rangle_\downarrow\right)\right],$$

(3)

which implies the familiar cosine-like correlations in the Bell setting (although written in an unusual form, these are the ordinary 30 possible combinations of Alice and Bob). It is, therefore, the measurement of one particle's spin orientation that nonlocally affects the orientation of the other, as the spin values are already fixed. But then, again, by the experiment's time evolution, the spin orientations must have emerged prior to the spin values (Fig. 4).

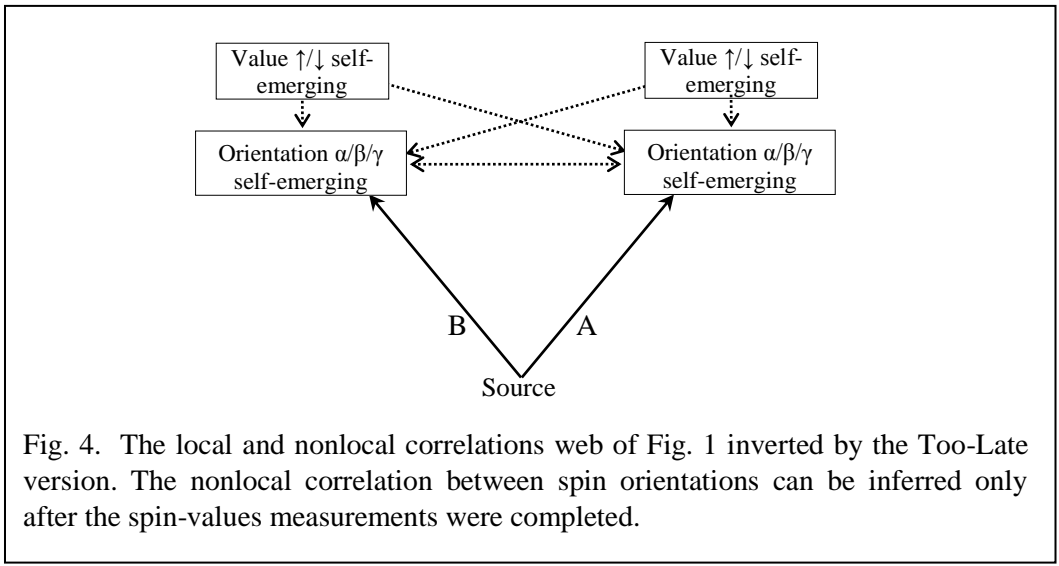

Fig. 4. The local and nonlocal correlations web of Fig. 1 inverted by the Too-Late version. The nonlocal correlation between spin orientations can be inferred only after the spin-values measurements were completed.

## 7. How Rigorous is the Argument? A Bohmian Account Countered as a Generic Alternative

It is now time for a more critical discussion. The core of our argument concerns the correlations between the early "choices" of SGM orientations, $α/β/γ$, taken by each of the EPR particles. These choices emerge *after* appropriately grouped by the later spin-value measurements, turning out to be strictly correlated. The crucial question is therefore: Why regard these correlations as retrocausally *imposed* rather than merely *revealed* by the later measurements?

Our argument's apparent weakness on this point lies in its reliance on Bell's proof, which, originally devised for non-locality, cannot be extended to retrocausality. The reason is simple: In the EPR setting, two spin orientations *e.g.*, $α$ and $β$, are deliberately *imposed* by the external agents Alice and Bob, forcing the spin values to behave nonlocally so as to comply with spin conservation. Our setting, in contrast, cannot do so with spin values, *e.g.*, impose two ↑'s. Consequently, Bell's celebrated banishment of local hidden variables does not rule out hidden variables that are *nonlocal but orthocausal*.

Fortunately, another control employed by our setting enables making this last resort highly strained. In what follows we focus on the most notable realistic alternative to retrocausality, namely Bohmian mechanics.

In the Too-Late Choice experiment, so the Bohmian account would go, each EPR atom (*i.e.*, the corpuscular part of the wave-function) enters only one SGM, and proceeds into only one of that SGM's arms. It is only its accompanying *guide wave* that splits and goes through all six paths. This way, each atom aided by its guide-wave, can inform the other atom about its spin orientation and value *long before the final detections*. This can explain even interference effects of the atom's three SGM paths, should one wish to prove that the atom has gone through all of them.

The Bohmian account, however, applies in a simple manner only to single particles. With more than one, its resorts to an abstract multidimensional configuration space leads to a much less realistic content. In our setting, this resort must be taken when the superposed atom absorbs the photon (Sec. 5). This is a unique situation: We have a superposed atom which is *excited*. Hence, if sufficiently delayed within the three SGMs, it will reemit the photon while remaining superposed.

Now, should we place three photon detectors next to the three magnets, one of them will absorb the photon, thereby "collapsing" the atom's position as well. We can, however, place a single detector far enough, equidistant from the three magnets, such that the photon's detection does not distinguish between the paths. This is essentially an interference effect, leaving the photon's origin superposed.[3] Once this procedure is performed on both atoms, the entanglement between them swaps to an entanglement between their photons. Consequently, nonlocal correlations between the photons' polarizations are expected (Fig. 5).

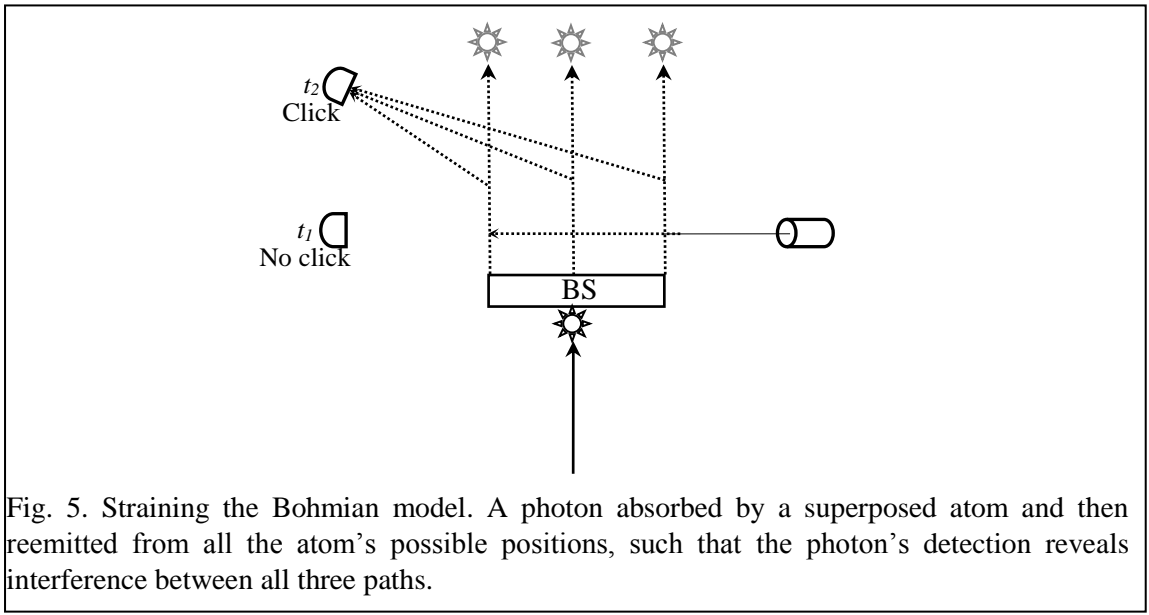

Fig. 5. Straining the Bohmian model. A photon absorbed by a superposed atom and then reemitted from all the atom's possible positions, such that the photon's detection reveals interference between all three paths.

Obviously the guide-wave model is highly strained at this point. *How can an empty wave absorb a photon (or worse, the photon's guide wave) and then keep carrying it in strict accordance with the atom's internal structure (electron shells, etc.) and finally emit it back, just like a corpuscular atom?*

In reply, Bohmian mechanics would argue that, when dealing with entangled particles, they no longer have well-defined and correlated positions in ordinary space but in *configuration space*. This resort to the mathematical realm seems to rob Bohmian mechanics of much of its realistic aspirations. So much so, even to the extent of strongly resembling its Copenhagen archrival, that one may wonder whether the medicine is not worse than the disease itself.

---

[3] It also resembles the Hong-Ou-Mandel effect [24], which can be seen as the quantum version of the Hanbury-Brown-Twiss experiment

This gedanken experiment's outcome, echoing similar sentiments in recent reviews, is part of a more detailed critique of Bohmian mechanics [25], to be published.

## 8. Summary: Why Go Time-Symmetric?

As noted earlier (Sec. 1), the merits of the time-symmetric interpretations have long ago been illustrated on the EPR experiment. The latter famously intrigues us by Alice's choice of spin orientation somehow affecting Bob's outcome and *vice versa*. This feat seems to be best elucidated if each measurement determines not only the particle's present state but its *earlier history* as well. Via the resulting spacetime zigzag, "nonlocal" turns out to be local in four dimensions (see also [7]).

But this account does more than mitigate the conflict between QM and special relativity. It also rids the particle from several other daunting tasks obliged by hidden-variables models. How, for example, can the presumed effect travel from one EPR particle to the other without being attenuated by distance? And how does this effect precisely locate the particle's twin, no matter where it went, how far, among a myriad of identical particles? The mathematical abstraction of Hilbert spaces addresses these questions, but, having no concrete counterpart in physical spacetime, it leaves them conceptually disturbing. The transactional account, in contrast, handles these questions just as easily as the nonlocality issue [5, 17, 18].

Other reasons for opting for the time-symmetric interpretations are also discussed in our critical examination [25] of the prevailing interpretations of QM. These can in divided into two major groups: *i*) abandonment of ontology *vs*. *ii*) excessive ontology. In the first group there are the Copenhagen school [26], QBism [27] and their like, which eschew the very notion of objective reality and admit only data, information, "knowledge," *etc.*, as ingredients of the quantum formalism. Consequently, no matter how unique and surprising a new effect predicted by QM can be, under this "shut up and calculate" spirit they are stripped of any mystery. Physics is thus denied the motivation to account for apparent anomalies, resolve paradoxes and search for better models. In the second group (*ii*), schools like Many Worlds [28] and Bohmian Mechanics [29] amend physical reality with entities that are *in principle* unobservable. For example, the parallel universes, numbering as many as all the possible outcomes of each and every quantum interaction anywhere in this universe, are inaccessible by definition. Similarly for Bohm's hidden variables: They must remain undetectable, otherwise violations of the uncertainty principle and SR would inevitably ensue

("hidden variables must be hidden forever" [30]). Moreover, in Sec. 7 we gave an example where, when faced with more than one particle, Bohmian mechanics assumes such an abstract form that amounts to "defection" to group (*i*).

Where do the time-symmetric interpretations belong according to this classification? They are certainly ontological, as they argue that spacetime objectively possesses some yet-unfamiliar properties. Yet this ontology is by no means excessive, neither untestable. If, at the quantum realm, causal effects proceed on both time directions, then sufficiently delicate experiments should be able to reveal this dual nature. Indeed TSVF already boasts some verifications of this kind [8], and further surprising theoretical and empirical results can be expected.

This of course comes with a price. If retrocausality underlies quantum phenomena, something very profound about time is still unknown, absent even in the revolutionary accounts of general relativity and quantum field theory. After a century of pondering quantum nonlocality and retrocausality, several authors have begun answering these very questions from various perspectives [31-34], indicating that this price may be fair.

**Acknowledgements**

It is a pleasure to thank Yakir Aharonov and William Mark Stuckey for very helpful comments and discussions. E.C. acknowledges support from ERC-AD NLST.